\documentstyle[12pt,psfig]{article}
\newcommand {\ignore}[1]{}
\def\lf{\leaders\hbox to 1em{\hss.\hss}\hfill}
\def\21{$SU(2) \ot U(1)$}
\def\321{$SU(3) \ot SU(2) \ot U(1)$}
\def\ne{\hbox{$\nu_e$ }}

\def\ns{\hbox{$\nu_{s}$ }}

\def\eq#1{{eq. (\ref{#1})}}

\def\VEV#1{\left\langle #1\right\rangle}

\def\lsim{\raise0.3ex\hbox{$\;<$\kern-0.75em\raise-1.1ex
\hbox{$\sim\;$}}}
\def\gsim{\raise0.3ex\hbox{$\;>$\kern-0.75em\raise-1.1ex
\hbox{$\sim\;$}}}

\def\beq{\begin{equation}}
\def\eeq{\end{equation}}
\def\bef{\begin{figure}}
\def\eef{\end{figure}}
\def\bet{\begin{table}}
\def\eet{\end{table}}
\def\bea{\begin{eqnarray}}
\def\eea{\end{eqnarray}}
\def\ba{\begin{array}}
\def\ea{\end{array}}
\def\bi{\begin{itemize}}
\def\ei{\end{itemize}}
\def\ben{\begin{enumerate}}
\def\een{\end{enumerate}}
\def\ra{\rightarrow}
\def\ot{\otimes}

\def\apj#1#2#3{          { Astrophys. J. }{\bf #1}, #3 (19#2)}

\def\aa#1#2#3{          { Astron. \& Astrophys.  }{\bf #1}, #3 (19#2)}

\def\ib#1#2#3{           { ibid. }{\bf #1}, #3 (19#2)}

\def\nps#1#2#3{        { Nucl. Phys. B (Proc. Suppl.) }{\bf #1}, #3 (19#2)} 
\def\np#1#2#3{           { Nucl. Phys. }{\bf #1}, #3 (19#2)}
\def\pl#1#2#3{           { Phys. Lett. }{\bf #1}, #3 (19#2)}
\def\pr#1#2#3{           { Phys. Rev. }{\bf #1}, #3 (19#2)}

\def\prl#1#2#3{          { Phys. Rev. Lett. }{\bf #1}, #3 (19#2)}

\def\zp#1#2#3{           { Zeit. fur Physik }{\bf #1}, #3 (19#2)}

\def\n.c.#1#2#3{         { Nuovo Cim. }{\bf #1}, #3 (19#2)}
\def\r.n.c.#1#2#3{       { Riv. del Nuovo Cim. }{\bf #1}, #3 (19#2)}
\def\sjnp#1#2#3{         { Sov. J. Nucl. Phys. }{\bf #1}, #3 (19#2)}

\renewcommand{\thefootnote}{\fnsymbol{footnote}}
\def\dfrac#1#2{{\displaystyle\frac{#1}{#2}}}

\begin{document}
\thispagestyle{empty}
\begin{titlepage}
\begin{center}
\rightline{hep-ph/9702372}
\hfill FTUV/96-81\\
\hfill IFIC/96-100\\
\hfill February 1997\\
\vskip 0.2cm
{\Large \bf Supernova Bounds on \\ 
Resonant Active-Sterile Neutrino Conversions\\
}
\vskip 1cm
{H.  Nunokawa}$^1$ 
\footnote{
E-mail: nunokawa@flamenco.ific.uv.es},
J. T. Peltoniemi $^2$ 
\footnote{
E-mail: Juha.Peltoniemi@Helsinki.fi},
A. Rossi$^1$
\footnote{
E-mail: rossi@gtae2.ist.utl.pt; $\,\,$ 
present address: Dept. de Fisica, Inst. 
Superior Tecnico, 1096 Lisbon Codex, Portugal},
and 
J. W. F. Valle$^1$
\footnote{
E-mail: valle@flamenco.ific.uv.es}\\
{\sl $^1$Instituto de F\'{\i}sica Corpuscular - C.S.I.C.\\
Departament de F\'{\i}sica Te\`orica, Universitat de Val\`encia\\
46100 Burjassot, Val\`encia, SPAIN\\
URL http://neutrinos.uv.es}\\
\vskip .2cm
{\sl $^2$ 
Department of Physics, Box 9, 00014 \\University of Helsinki, FINLAND
}

\vskip .2cm
{\bf Abstract}
\end{center}
\begin{quotation}

We discuss the effects of resonant $\nu_e \rightarrow\nu_s$ and 
$\bar{\nu}_e\rightarrow\bar{\nu}_s$ ($\nu_s$ is a {\it sterile} neutrino)
conversions  in the dense medium of a supernova. In particular, we assume 
the sterile neutrino $\nu_s$ to be in the hot dark matter few eV mass range.
The implications of such a scenario for the supernova shock re-heating, 
the detected $\bar\nu_e$ signal from SN1987A and for the $r$-process 
nucleosynthesis hypothesis are analysed in some detail. The resulting
constraints on mixing and mass difference for the $\nu_e-\nu_s$ system 
are derived. There is also an allowed region in the neutrino parameter 
space for which the $r$-process nucleosynthesis can be enhanced. 

\end{quotation}
\end{titlepage}

\renewcommand{\thefootnote}{\arabic{footnote}}
\setcounter{footnote}{0}
\section{Introduction}

The possibility that light sterile neutrinos can be mixed in 
the leptonic charged current seems to be the very appealing from 
the point of view of the present {\it anomalies} observed in the 
neutrino sector:  the solar \cite{SNP} and atmospheric \cite{ANP}
neutrino problems  as well as the need for a few eV mass 
neutrino as the hot dark matter in the Universe \cite{SS1,PHKC}. 
Barring the possibility that the three active neutrinos 
are nearly degenerate in mass \cite{DEG}, the simplest way to
simultaneously account for these observations is to postulate 
\footnote{The measurement of the $Z$ boson width at LEP 
has provided a limit on the number of active neutrino types  
with mass smaller than $M_Z/2$, $N_\nu = 3.09\pm0.13$ \cite{PDG}. 
However, it cannot preclude the existence of light sterile neutrinos.} 
the existence of a light sterile neutrino \cite{PTV,PV,all}. 
Moreover, some of these scenarios may also account for neutrino 
oscillations between the electron neutrino and  muon neutrino, 
as possibly hinted at the LSND experiment \cite{lsnd}.

The conversions between neutrino species may change 
significantly the phenomena occurring in supernova.
Of particular interest is the conversion to a sterile
neutrino ($SU(2)_L$ singlet), since the sterile state does 
not interact at all with the supernova matter. 

Here we focus on the resonant conversions of electron neutrinos
(or anti-neutrinos) to sterile neutrinos outside the neutrinosphere.
For mass difference $\delta m^2 < 10^4  \mbox{eV}^2 $ the MSW 
resonance will occur in these regions of the supernova where 
neutrinos  freely stream.
Note that for this mass range the conversions to sterile 
neutrinos in the inner core can be neglected, for all values of
the mixing angles \cite{KMP}. Hence, in the absence of other 
non-standard interactions, such as a large magnetic moment,
the sterile neutrinos are not emitted from the supernova core. 
We will work always under this assumption.
Moreover, we assume that the mass eigenstate consisting mainly of 
the singlet state is heavier, with mass between 1 eV and 100 eV. 
The opposite case would require however special forms of 
the mass matrices in order to avoid the constraints from 
neutrino-less double beta decay. This possibility certainly
exists but, for definiteness we do not consider.
Finally, we assume that no other neutrino conversions, such
as flavour conversions, take place. Generalization to the case 
of three-neutrino flavours is straightforward but model-dependent. 

The effects of such active to sterile transitions have been 
previously discussed in \cite{MS1,SS,Raffelt}, mainly concentrating 
on the effects for neutrino observations at terrestrial detectors.
In the present paper we consider the implications of resonant 
$\nu_e\rightarrow\nu_s$ or $\bar{\nu}_e\rightarrow\bar{\nu}_s$ 
conversions in the dense supernova medium on the neutrino re-heating 
of the shock wave \cite{CW,Wilson85}, on the $\bar\nu_e$ signal in 
 terrestrial detectors, as well as on supernova heavy-element 
nucleosynthesis \cite{Woosley}. 

The effect of the electron neutrino to sterile neutrino 
conversion on supernova nucleosynthesis (via rapid neutron 
capture process or so-called $r$-process) has not yet been discussed 
with sufficient detail. So far most studies have concentrated on the 
case of active neutrinos, giving stringent limits on electron, muon and 
tau neutrino conversions \cite{QF,massless}. Nevertheless, it has been 
suggested \cite{juha2} that a resonant conversion from muon neutrinos 
to sterile neutrinos, with a subsequent conversion between electron 
and muon neutrinos, would enhance $r$-process nucleosynthesis. The 
required conversion pattern appears naturally in some specific 
models \cite{PTV}.

Neutrino conversions may also influence the explosion mechanisms
of the supernova in different ways.  Although the main interest
lays in the transitions among the active neutrino species
\cite{Fu}, it has also been suggested that conversions between 
sterile and active neutrinos may enhance the explosion 
\cite{juha1}. However, for the mass range we are considering,
such an enhancement is not possible, the conversions weaken the 
shock wave, preventing the explosion.

In Sec. 2  we give a quick reminder on the picture of the neutrino 
propagation in matter and on the resonant $\nu_e -\nu_s$ conversion 
\cite{MS}.
In Sec. 3 we give a qualitative discussion of the electron 
concentration $Y_e$ in supernova and on the effects of the 
active-sterile neutrino conversions on $Y_e$ and, in turn, 
on the neutrino evolution itself. This can lead to non-trivial 
feedback effects \cite{feedback0,feedback1,feedback2}.
Sec. 4.1 discusses the implications of $\nu_e -\nu_s$ conversions 
for the neutrino re-heating mechanism.  In Sec. 4.2 we analyse 
the impact of our scenario in the later epoch of supernova 
evolution (few seconds after the core bounce) for the 
supernova (anti)neutrino detection (Sec. 4.2.1) as well as for 
$r$-process nucleosynthesis  (Sec. 4.2.2). In Sec. 5 we summarize 
our results and discuss their significance, by comparing the supernova
limits we derive with the laboratory and nucleosynthesis limits
on $\nu_e -\nu_s$ conversions.
 
\section{The active-sterile neutrino resonant conversion}

In our discussion we only consider the conversion channels 
$\nu_e\rightarrow\nu_s$ and $\bar{\nu}_e\rightarrow\bar{\nu}_s$ 
where \ns ($\bar{\nu}_s$)  is the sterile neutrino
\footnote{In the ultra-relativistic limit $\nu_s$ and $\bar{\nu}_s$ 
have opposite helicity.}.
For the sake of simplicity we will not consider the effect of 
$\nu_e\leftrightarrow\nu_{\mu,\tau}$ and 
$\bar{\nu}_e\leftrightarrow\bar{\nu}_{\mu,\tau}$ conversions.  
In the following we consider the $\delta m^2 = m^2_2 -m^2_1 > 0$ case, 
corresponding (for sufficiently small mixing angle) to the situation 
in which the heavier state is mostly the sterile neutrino. 

The evolution of the $\nu_e -\nu_s$  system in the matter background 
is determined by the Schr\"oedinger equation
\begin{eqnarray}
i{\mbox{d} \over \mbox{d}r}\left(\matrix{
\nu_e \cr\ \nu_s\cr }\right) & = & 
 \left(\matrix{
 {H}_{e}
& {H}_{es} \cr
 {H}_{es} 
& {H}_{s} \cr}
\right)
\left(\matrix{
\nu_e \cr\ \nu_s \cr}\right) \,\,, \\
  H_e & \! = &  \! 
 V_e - \frac{\delta m^2}{4E_\nu} \cos2 \theta\,, \,\,\,\,\,\,\,\,\,
H_s \!= V_s+  \frac{\delta m^2}{4E_\nu} \cos2 \theta, \nonumber \\
H_{es}& \!= &  \frac{\delta m^2}{4E_\nu} \sin2 \theta, \\
\nonumber
\label{evolution1}
\end{eqnarray}
where the effective potential $V_e$ for $\nu_e $ arises from 
the coherent forward neutrino scattering off-matter constituents 
\cite{MS} and is given by 
\footnote{The effective potential should also contain
contributions from the neutrino background. 
We have ignored them since the neutrino densities in the relevant 
regions are at least one order of magnitude smaller than the 
corresponding electron densities, and the neutrino terms in 
the effective potential involve an additional suppression
factor because most neutrinos travel almost in the same
direction.}
\begin{eqnarray}
\label{potential}
V_e & = &\frac{\sqrt{2}G_F \rho}{m_N} (Y_e- \frac{1}{2}Y_n)=
\frac{\sqrt{2}G_F \rho}{2m_N} (3Y_e- 1)\,, 
\\
Y_e& \equiv & \frac{n_e}{n_e+n_n} \, ,\hskip 1 cm Y_n = 1-Y_e\, .\nonumber
\end{eqnarray}
Here $G_F$ is the Fermi constant, $\rho$ is the matter density,  $m_N$ is 
the nucleon mass and  $n_e$ and $n_n$ are the net electron and the neutron 
number densities in matter, respectively. Note that charge neutrality 
$Y_p=Y_e$ is assumed and that there is no potential for $\nu_s$, i.e., 
$V_s =0$.
For the system  $\bar{\nu}_e\rightarrow\bar{\nu}_s$  the matter potentials
just change their sign.

The resonance condition reads as:
\begin{equation}
V_e = \frac{\delta m^2}{2E_\nu} \cos2\theta ~.
\label{rc}
\end{equation}

Let us remind that for $\delta m^2 >0$, either the conversion 
$\nu_e\rightarrow\nu_s$ (for $V_e > 0$ i.e. $Y_e> 1/3$) or 
$\bar{\nu}_e\rightarrow\bar{\nu}_s$ (for $V_e <0$ i.e. $Y_e < 1/3$) 
takes place. This is important because, as we will see later, in the 
region above the neutrinosphere the matter potential $V_e$ changes 
its sign due to the different chemical content. For our later 
discussion, it is instructive to know the profiles of the 
matter density and of the electron fraction $Y_e $ 
outside the neutrinosphere.
In Fig. 1a and 1b we plot these quantities for $t< 1$ s  
post-bounce (in short p.b.) and $t> 1$ s p.b. as given by 
the Wilson supernova model \cite{qian}. We can see that the 
electron concentration $Y_e$ is rather low just near the neutrinosphere,  
$Y_e \approx 0.1$ and 0.01 for the earlier and later epoch, 
respectively and far away it increases to values $\gsim$ 0.4.  
On the other hand, the matter density $\rho$ exhibits a 
monotonically decreasing behaviour. 
In Fig. 2a and 2b we plot the modulus of the effective matter 
potential $V_e$ using the matter density and $Y_e$ profiles as 
given in Fig. 1a and 1b.  The position where $Y_e$ takes the
value $1/3$ (i.e. $V_e=0$) is indicated by $r^*$. This position 
corresponds to $r^* \approx 160 $ km and 12 km for the earlier 
and later epochs, respectively.  Clearly, the effective potential 
$V_e$ changes its sign from negative to positive at the point $r^*$. 

The resonance condition in the \eq{rc} provides the $\delta m^2$ 
value for which neutrinos with some given energy can experience 
the resonance for a certain value of the potential (or equivalently,  
at some position $r$). For the sake of convenience, in the right 
ordinate of Fig. 2 we have also indicated such  corresponding values 
of $\delta m^2$ for typical neutrino energy  $E= 10$ MeV. 
We see that for $\delta m^2 \gsim 10^{2}$ eV$^2$ only
$\bar{\nu}_e \rightarrow \bar{\nu}_s$ conversions can take place, 
and this happens in the region where  $Y_e \leq 1/3$. 
On the other hand for smaller values, $\delta m^2 \lsim 10^{2}$ eV$^2$ 
there can occur three resonances. The $\bar\nu_e$'s are first converted, 
say at $r_1 < r^*$, then there are two resonance points at $r_2$ and $r_3$,  
where $r_3> r_2 >r^*$ (i.e. in the region where $Y_e > 1/3$), for the 
${\nu}_e \leftrightarrow {\nu}_s$ channel. In order to illustrate 
this more explicitly we plot in Fig. 3 the schematic level crossing 
diagram for ${\nu}_e-{\nu}_s$ and $\bar{\nu}_e-\bar{\nu}_s$ system,
assuming the mixing angle to be small. From this figure, it is also 
manifest that the $\nu_s$'s originated from the first  $\nu_e$ 
conversion  (at $r_2$) can be re-converted into $\nu_e$'s at 
the second resonance (at $r_3$). 
In our subsequent discussion, we will employ the simple Landau-Zener 
approximation \cite{Landau,HPD} to estimate the survival probability 
after the neutrinos cross the resonance. Under this approximation, 
the $\nu_e$ (or $\bar\nu_e$) survival probability is given by 
(in the case of small mixing angle)
\begin{eqnarray}
\label{LZ}
 P & = & 
\exp\Biggl(-\dfrac{\pi^2}{2}\dfrac{\delta r}
{L_{m}^{\rm res}} \Biggr) \nonumber \\
          & \approx & \exp\left[
-2 \times  10^{-5} \times \sin^22\theta 
\left(\dfrac{\delta m^2 }{\mbox{eV}^2} \right)^2
\left(\dfrac{10\mbox{MeV}}{E_\nu}
\right)^2 \left(
\dfrac{\mbox{d}V_e}{\mbox{d}r}  \times 
\dfrac{\mbox{km}}{\mbox{eV}}
\right)^{-1}_{res}
                 \right] \ , 
\end{eqnarray}
where $L_{m}^{\rm res}$ is the neutrino oscillation length at 
resonance.  
Notice that for $\delta r /L_{m}^{\rm res}>1$ the 
resonant neutrino conversion will be adiabatic \cite{MS}. 
We can expect the maximal sensitivity to the mixing angle for 
$\delta m^2 =10^4$eV$^2$. From Fig. 2a and 2b one can estimate
that in this case the resonance occurs at high density 
$\rho \sim 10^{11}-10^{13}$g cm$^{-3}$ just above the 
neutrinosphere, where the gradient $\dfrac{\mbox{d}V_e}{\mbox{d}r} \sim 
1- 10^{-2}$eV km$^{-1}$. From the \eq{LZ} we can estimate that the 
conversion will be adiabatic for $\sin^2 2 \theta\gsim 10^{-6}- 10^{-4}$.

Due to the double resonances for the $\nu_e \leftrightarrow \nu_s$ channel, 
the $\nu_e$ survival probability after the second resonance is given by 
\begin{equation}
\label{prob}
P(\nu_e \rightarrow \nu_e ) = P(r_2) P(r_3) + [1-P(r_2)] [1-P(r_3)],
\end{equation}
where $P(r_2)$ and $P(r_3)$ are the survival  probabilities 
calculated according to the \eq{LZ} at $r_2$ and $r_3$, respectively. 

\section{The feedback induced by the neutrino conversion}

All the neutrino species emitted from the neutrinosphere
have approximately the same luminosity $L_\nu$ after a few
ms p.b., characterised by a thermal Fermi distribution with 
temperature $T_\nu$  and zero chemical potential.
The typical duration of the neutrino emission is about 10 seconds.

\subsection{Neutrino emission and absorption reactions and 
$Y_e$ profile in supernova}

In the region outside the neutrinosphere, due to the intense 
neutrino radiation, the electron fraction is determined by the 
neutrino capture by nucleons and by their reverse processes:
\begin{eqnarray}
\label{nu-n}
\nu_e+n &\leftrightarrow &p+e^- ~,\\
\label{nu-p}
\bar\nu_e+p &\leftrightarrow &n+e^+ ~ .
\end{eqnarray}
In particular $Y_e$ above the neutrinosphere is approximately 
given by \cite{QF}
\begin{equation}
\label{ye}
Y_e \approx \dfrac{ \lambda_{e^+ n} +  \lambda_{\nu_e n} }
{ \lambda_{e^- p} +  \lambda_{e^+ n} +  
\lambda_{\bar\nu_e p}+ \lambda_{\nu_e n}  }~.
\end{equation}
The neutrino capture rates  depend essentially on the neutrino 
luminosity $L_\nu$ and energy $E_\nu$,  
\begin{equation}
\label{rates}
\lambda_{\nu N}\approx 
\int_0^\infty \hskip -0.3cm \phi^0(E_\nu)
\sigma_{\nu  N} (E_\nu)dE_\nu 
\propto {L_\nu\over \langle E_\nu\rangle}\langle E_\nu^2 \rangle
\propto L_\nu \langle E_\nu \rangle \, ,
\end{equation}
where $(\nu,\ N)=(\nu_e,\ n)$ or $(\bar\nu_e,\ p)$.  
The neutrino luminosity is given by the black-body surface emission formula, 
$L_\nu = 5.6 \times  10^{46}  r_\nu^2 ~ T_\nu^4$ ergs/s 
(here $r_\nu$ is given in km and $T_\nu$ in MeV).
For our purpose it is sufficiently accurate to assume the 
neutrino differential flux $\phi^0(E_\nu)$ to be given by

\begin{equation}
\label{f0}
\phi^0(E_\nu) = 
\frac{L_\nu}{4\pi r^2}\dfrac{ E^2_\nu / (\mbox{e}^{E_\nu/T_\nu}+1)} 
{ \int E^3_\nu \mbox{d}E_{\nu} / (\mbox{e}^{E_\nu/T_\nu} +1) } ~~.
\end{equation}
On the other hand, the rates for the inverse processes depend 
strongly on the matter temperature $T$ and are given as 
\begin{equation}
\label{lambdaback}
\lambda_{e N} \approx \frac{1}{\pi^2}
\int_0^\infty \hskip -0.3cm 
\frac{\displaystyle \sigma_{e N}(E_e){E^2}_e}
{\exp[(E_e\mp\mu_e)/T] + 1}    dE_e \propto  T^5  \ ,
\end{equation}
where $\sigma_{e N}$ is the electron or positron 
capture cross section with $(e,\ N)=(e^-,\ p)$ or $(e^+,\ n)$. 
(In the above formula the negative sign in front of $\mu_e$ 
is for electrons and the positive sign is for positrons.)

In Fig 4a and 4b we have plotted the above rates 
$\lambda_{\nu N}, \lambda_{e N}$ (for the earlier and 
later epochs, respectively) as a function of the distance 
from the centre of the star. For convenience we have also plotted 
the matter temperature profile by the solid line (see the 
right ordinate scale). 
One can see from Fig. 4a and 4b that close to the neutrinosphere
the rate $\lambda_{e^- p}$ dominates over the others, and hence $Y_e$ 
is less than 1/3 (see the \eq{ye}), as can be seen from Fig. 1a and 1b. 
As $r$ increases $\lambda_{e^- p}$ decreases faster than $\lambda_{\nu N}$ 
and hence $Y_e$ increases. At some point $r\sim r^*$ where
$\lambda_{e^- p} \sim \lambda_{\nu N}$, the fraction $Y_e$ takes the 
value $1/3$. 
In the region $r>r^*$ where the neutrino absorption reaction is high enough 
to dominate the proton-to-neutron ratio, the electron fraction is given by
\begin{equation}
\label{yelater}
Y_e\approx {\lambda_{\nu_en}\over\lambda_{\bar\nu_ep}+\lambda_{\nu_en}}
\approx {1\over 1+\langle E_{\bar\nu_e}\rangle/\langle E_{\nu_e}\rangle}.
\end{equation}
assuming the fluxes to be given as in the \eq{f0}. 
For the typical energies $\langle E_{\nu_e} \rangle \sim 11$ MeV
and  $\langle E_{\bar{\nu}_e} \rangle \sim 16$ MeV at $t> 1$ s p.b., 
 we find $Y_e = 0.41$, in good agreement with numerical supernova
models \cite{QF}. The above discussion explains well the 
behaviour of the $Y_e$ profiles shown in Fig. 1a and 1b. 

\subsection{Feedback effect on neutrino conversion}

Now let us come to the main point concerning the effect of the 
neutrino conversion upon the electron content in the matter. 
Suppose that at some position $r_0$ and time $t_0$ neutrino 
conversions $\nu_e \rightarrow \nu_s$ or 
$\bar{\nu}_e\rightarrow\bar{\nu}_s$ occur. 
{}From the \eq{rc} and \eq{LZ} we see that the conversion probability 
depends on the value of $Y_e$ and its derivative at the resonance 
position $r_0$. The neutrino conversions $\nu_e \rightarrow \nu_s$ or 
$\bar{\nu}_e\rightarrow\bar{\nu}_s$ reduce the value of 
$\lambda_{\nu_e n}$ or $\lambda_{\bar\nu_e p}$ and this could lead 
to the modification of $Y_e$ in the region $r>r_0$ (see the \eq{ye}). 
If the modification of $Y_e$ is fast enough it could affect 
any subsequent neutrino conversion  occurring at some position 
$r>r_0$ and time $t> t_0$. 
This subsequent neutrino conversion could again modify 
the value of $Y_e$, and so on. In this way, neutrino 
conversion rates and the electron fraction continuously 
affect each other during the neutrino emission, leading 
to non-trivial feedback phenomena for the neutrino 
transitions. These should in principle be taken into
account self-consistently in neutrino conversion studies. 
It is well known that in the early universe feedback effects 
on the neutrino background should be reflected in the evolution 
of neutrinos \cite{feedback0} and such effects have been 
studied in ref. \cite{feedback1}. On the other hand, 
in the context of supernova physics, feedback effects 
on the neutrino background have recently been studied in ref. 
\cite{feedback2}. In what follows we will consider the 
analogous effect for the electron abundance $Y_e$.

{}From our discussion in Sec. 3.1, we can immediately understand 
that the feedback is operative only when 
$\lambda_{\nu N} \gg  \lambda_{eN} $ (see the \eq{ye}). 
Right after the bounce ($t<1$ s p.b.) the relevant resonance 
layers lie so far away from the stellar core that the 
$\lambda_{\nu N}$ is too small
since $\phi^0 \propto r^2_\nu/r^2$ (cfr. \eq{f0} ) and as a 
result feedback effects would be small. 
On the other hand, later on ($t>1$ s p.b.) the neutrinosphere 
shrinks so much that $\lambda_{\nu N}$ is high at the relevant 
resonance positions (now they lie much  closer to the centre) 
and, as a result, the feedback effects would be potentially
important. 

The neutrino conversion could be affected by the feedback 
either by shifting of the resonance position, or through
a change in the adiabaticity.
Let us first see the effect on the resonance position. For 
simplicity let us assume that $Y_e$ is determined only by 
$\lambda_{\nu N}$ as in \eq{yelater}. In the region where 
$Y_e > 1/3$ ($Y_e < 1/3$) the $\nu_e\rightarrow\nu_s$ 
($\bar{\nu}_e\rightarrow\bar{\nu}_s$) conversions occur 
and this tends to decrease (increase) $Y_e$ in the region 
above the resonance layer because $\lambda_{\nu_e n}$ 
($\lambda_{\bar\nu_e p}$) decreases. 
This could affect the resonance condition for the subsequent 
neutrinos, leading to a shift of the resonance position.  
Moreover, if the value of $Y_e$ becomes less or equal to
1/3, the subsequent $\nu_e\rightarrow\nu_s$ conversion would 
be suppressed because the resonance condition is not satisfied. 
This is therefore a negative feedback. 
Analogously, a   negative feedback is also 
present for the $\bar{\nu}_e\rightarrow\bar{\nu}_s$
channel.

Let us now turn to the feedback effect on the adiabaticity. 
The neutrino conversion would change the potential gradient 
gradually, as neutrinos 
of different parts of the energy spectra reach the resonance.
In the case of $\bar{\nu}_e\rightarrow\bar{\nu}_s$ conversion,  
the depletion of the anti-neutrino flux drives $Y_e$ up by 
decreasing the anti-neutrino capture rate, and so it tends 
to raise the absolute value of the gradient of the potential. 
This weakens the adiabaticity of the neutrino conversion. 
As for the $\nu_e \rightarrow \nu_s$ conversion, in the
increasing part of the potential in Fig. 2, due to the decrease 
of $Y_e$ caused by \ne to \ns conversion, the shape of the potential 
would be flattened, leading to better adiabaticity. On the other hand, 
for the decreasing part of the potential (see again Fig. 2), the potential 
gradient would be steepened leading to worse adiabaticity. 

Let us now consider the relevance of the feedback in terms of 
the $\delta m^2$ involved.
We can distinguish three different 
ranges of $\delta m^2$: (i) $10^3$ eV$^2 < \delta m^2< 10^4$ eV$^2$,
 (ii) $10^2$ eV$^2 < \delta m^2< 10^3$ eV$^2$ and  
(iii) $\delta m^2< 10^2$ eV$^2$. 
For the mass range (i) and (ii) (see Fig. 2), only anti-neutrino 
conversions take place, while for the range (iii) both neutrino 
and anti-neutrino conversions take place. 
In range (i), the transitions occur close enough to the neutrino 
sphere, where the $e^- p \rightarrow \nu_e n$ reaction dominates over 
the corresponding neutrino absorption reactions. 
In this case $Y_e$ 
around the resonance position is not affected by the
$\bar{\nu}_e\rightarrow\bar{\nu}_s$ conversion. 
Hence the feedback is irrelevant for this mass range. 

On the other hand, for the range (ii), the conversions occur farther 
away from the neutrinosphere where the $e^- p \rightarrow \nu_e n$ 
rate becomes comparable to that of $\bar\nu_e p \rightarrow e^+ n$, 
especially in the later epoch.
Hence the anti-neutrino conversion would be somewhat more affected 
by the feedback than in the previous case. However, the effect is still
not large, since $\lambda_{e^- p}$ is not small compared to 
$\lambda_{\nu N}$. 

Finally, for the range (iii) the situation is more complicated. 
In this case three resonant conversions can occur:  one in 
the $\bar\nu_e\rightarrow \bar\nu_s$ channel and two in the 
$\nu_e \leftrightarrow\nu_s$ channel (see Fig. 3). 
For the $\bar\nu_e\rightarrow\bar\nu_s$ resonant conversion
(at $r_1 < r^*$) the preceding conclusion still holds and the 
feedback is not important. 
Similarly the next conversion $\nu_e\rightarrow\nu_s$ 
(at $r_2 > r^*$) is not too affected by the feedback due to
its proximity to the core, as before. 
Moreover, this $\nu_e\rightarrow\nu_s$  conversion
is expected to have the same degree of adiabaticity 
as that of the previous anti-neutrino conversion, due 
to roughly the same steepness of the matter potential.
As a result $P(\bar{\nu}_e\rightarrow\bar{\nu}_e ; r_1) \approx 
P(\nu_e\rightarrow\nu_e ;r_2)$
and, to some extent the change in $Y_e$ at $r_2$ will be 
compensated by that at $r_1$, leading to a small net feedback
\footnote{This observation is strictly true in the case 
of absence of the feedback. However, since the feedback 
effect may slightly decrease the adiabaticity of 
anti-neutrinos and increase the adiabaticity of 
neutrinos, the conversion probability for neutrinos
may be a little larger than for anti-neutrinos.}.

Now let us come to the last resonance in the channel 
$\nu_e\leftrightarrow\nu_s$, at $r_3$. This conversion 
could be significantly affected by the feedback because 
in the region around $r_3\ (\gg r^*$) $Y_e$ is determined 
only by the $\lambda_{\nu N}$ reaction rates, since the
inverse rates are small, as seen from Fig. 4b.
For the sake of discussion, let us first consider the case 
where the conversions at $r_1$ and $r_2$ are sufficiently adiabatic. 
In this case the resonance at $r_3$ is also expected to be 
very adiabatic, since the slope of the potential around $r_3$ 
is much less steep than that for the previous resonances. 
Therefore, the $\nu_e$ flux would be completely recovered by 
the last conversion $\nu_s \rightarrow \nu_e$ at $r_3$. 
This last conversion $\nu_s\rightarrow \nu_e$ would remain 
adiabatic because the increase of $Y_e$ due to the conversion 
would not violate the adiabaticity but simply shift the
resonance position. As a result we conclude that it
would not be affected by the feedback.
In the opposite case, when at $r_1$ and $r_2$ the conversions 
are not adiabatic ($P \sim 1$), the feedback effect, due to the last 
$\nu_e\leftrightarrow\nu_s$ transitions, could be very important.   
The depletion of the electron neutrino flux and the unchanged 
$\bar\nu_e$ flux would lower the value of $Y_e$ and hence $V_e$,  
with the effect of shifting  the  resonance position 
$r_3$ at lower values. 
At the same time the potential gradient becomes steeper
and hence the conversions become less adiabatic, leading to the 
suppression of the $\nu_e\rightarrow \nu_s$ transition. In this 
case we would have a net feedback effect on the neutrino conversion. 
This situation occurs for rather small mixing angle 
$\sin^2 2\theta < 10^{-2}$ and for $\delta m^2 < 10^2$ eV$^2$. 
We can have a quantitative insight of the significance of the effect 
from the  \eq{yelater} by simply requiring $Y_e \geq 1/3$. For 
simplicity, let us assume $P_{\bar\nu_e}(r_1) \sim  P_{\nu_e}(r_2)\sim 1$.  
Thus we can deduce a lower bound for the survival probability  at $r_3$, 
$P_{\nu_e}(r_3) \gsim 0.7$. 
In other words, the second $\nu_e\rightarrow \nu_s$ conversion 
would be stopped when 
about 30 \% of the lower part of $\nu_e$ spectra are 
converted to $\nu_s$ (because the lower energy neutrinos undergo 
resonance before the higher ones) even if the conversion is 
initially very adiabatic. 

Instead of taking into account this feedback effect upon the 
neutrino evolution, we have simply stop the $\nu_e\rightarrow\nu_s$ 
conversion when $Y_e$ reaches the value of 1/3. 

We conclude that the feedback effects would be small and could 
be neglected in the earlier epoch in the region relevant for 
our discussion. As for the later epoch, the feedback may be 
more relevant especially for the range $\delta m^2 <10^2$ eV$^2$.    

\section{Constraining Neutrino Parameters}

In this section we are mainly concerned with the
implications of active-sterile neutrino conversions for
supernova physics. For the earlier epoch of supernova
evolution active-sterile neutrino conversions would
suppress the shock re-heating.
For the later epoch  active-sterile neutrino conversions 
could suppress the detected $\bar\nu_e$ signal from SN1987A.   
On this basis we derive stringent constraints on the neutrino parameters. 
On the other hand, these conversions could affect the $r$-process 
nucleosynthesis in the later epoch, 
either to suppress it or to enhance. In the first 
case we again analyse the restrictions on neutrino parameters.

\subsection{Earlier Epoch: $t< 1$ second after core bounce}

In the following we consider only the epoch after the core bounce 
and the neutrino evolution in the regions outside its neutrinosphere. 
Indeed, for the range $\delta m^2 \leq 10^4$ eV$^2$ neutrino 
transitions in the dense matter of the core ($\rho \sim 
10^{13}$ g/cm$^3$) are strongly suppressed \cite{KMP,SS}.  

At this time a {\it reflected} shock wave is formed between 
the inner homologously collapsing part and the supersonically 
falling outer portion of the initial iron core. This shock 
propagates through the mantle and may result in mass ejection 
when it reaches the surface of the star. In fact the shock 
suffers energy loss due in particular to the emission 
of neutrinos which prevents a successful explosion of the star. 
As soon as the shock wave has passed through the neutrinosphere 
there is a large burst of $\nu_e$'s. Subsequently on a time scale 
of several seconds after the core bounce the emission of all neutrino 
species drives the evolution of the star to the final cool and 
neutron  star. The epoch within 1 second  after the core bounce  is rather 
important for the re-heating of the shock wave. 
In the delayed explosion mechanism \cite{CW,Wilson85} the 
neutrino energy deposition, occurring between the neutrino 
sphere and the site where the shock is stalled, can re-start 
the shock and  power the explosion. 
As can be seen from Fig. 4a, at $r\sim 300\div 400 $ km away from 
the neutrinosphere the neutrino absorption rate on free nucleons 
dominates over the capture of electrons. At this position the  
energy transfer from neutrinos to the matter takes place so
as to  help the shock
\footnote{Other mechanisms for energy deposition, like neutrino 
scattering off-electrons or neutrino-anti-neutrino annihilation 
are less efficient \cite{CVB} and thereby we neglect them 
in the following.}.  

In the absence of neutrino conversions the corresponding energy 
gain (per nucleon) $\frac{dE}{dt}\equiv \dot{E}(t)$ is 
\begin{equation}
\label{rate1}
 \dot{E}(t) 
\approx \frac{L_\nu \sigma_{\nu N} }{4\pi r^2} 
\end{equation}
where $L_\nu$ is the total $\nu_e + \bar{\nu}_e$  luminosity and 
$\sigma_{\nu N} \sim 9 \times  10^{-44}\times E_\nu^2$ cm$^2$. 
For $L_{\nu} \sim 3\times 10^{52}$ ergs/s, $E_{\nu} = 10$ MeV and 
$r= 300$ km, we find $ \dot{E}(t)  \sim 20$ MeV s$^{-1}$. 
This rate seems to be large enough on the time scale of 
0.1-0.2 second if compared with the gravitational 
potential (per unit mass) $G_N M_r/r\sim$ 7-10 MeV of the material  
stopped behind the shock ($M_r\sim 1.5 M_\odot$ is the included mass). 
Thus the  neutrino energy transfer can help the material to overcome 
the gravitation of the star and so to escape it.    
The success of this scenario strongly depends on the neutrino 
luminosity. Clearly the sterile conversion $\nu_e\rightarrow\nu_s$ or 
$\bar{\nu}_e\rightarrow\bar{\nu}_s$ would imply a depletion of active 
neutrino luminosity and thereby can spoil the re-heating process \cite{SS}. 

We have calculated the ratio $R$ of the neutrino heating rate
in the presence of $\nu_e \rightarrow\nu_s$ and 
$\bar\nu_e \rightarrow\bar\nu_s$ transitions to the
corresponding rate in the absence of such transitions.
Following ref. \cite{Wilson85} we use an approximate 
expression for $R$, in which we neglect the re-emission 
of neutrinos by the heated matter, leading to
\begin{eqnarray}
\label{ratio}
 R & = &\frac{ Y^{\prime}_n   \dot{E}^\prime_{\nu_e n}(t) + 
Y^\prime_p   \dot{E}^\prime_{\bar\nu_e p}(t) } 
{ Y_n   \dot{E}_{\nu_e n}(t) + 
Y_p   \dot{E}_{\bar\nu_e p}(t) } ~~, \\
   \dot{E}_{\nu N}(t) & \sim &
 \int E_\nu \sigma_{\nu N} \phi^0 (E_\nu)\mbox{d}E_\nu ~~, ~~~~~
\dot{E}^\prime_{\nu N}(t)  \sim 
 \int E_\nu \sigma_{\nu N} P(E_\nu)
 \phi^0(E_\nu)\mbox{d}E_\nu ~.  
\end{eqnarray}
where the primed quantities $Y^\prime_p$  and $Y^\prime_n$ 
stand for the proton and neutron abundances calculated in the
presence of active-sterile neutrino conversions.

In Fig. 5 we plot the iso-contour for different values of 
the ratio $R$ in the parameter space ($\delta m^2, \sin^2 2 \theta$). 
Requiring  a moderate effect $R > 0.9$,  
one can exclude $\sin^2 2\theta> 10^{-8}$ for  
$\delta m^2\sim10^4$ eV$^2$,  
whereas for smaller mass,  $\delta m^2 ~ 1 \div 10$ eV$^2$, the bound on 
the mixing is weaker and lies in the range 
$\sin^2 2\theta  > 7  \times  10^{-5}\div5 \times  10^{-3}$. Note that our
bounds are in qualitative agreement, though slightly more stringent
than those found e.g. in ref. \cite{SS}.

\subsection{Later Epoch: $t >1$ second after core bounce}

For the later epoch we consider the effect of active-sterile 
neutrino conversions both on the $\bar\nu_e$ signal as well as
the $r$-process nucleosynthesis and analyse the possible 
restrictions on neutrino parameters.

\subsubsection{Implications for the detection of SN1987A $\bar\nu_e$ signal}
 
The Kamiokande II and IMB detectors observed 11 and 8
$\bar\nu_e$ events, respectively, from SN1987A \cite{KA,IMB}. 
This is in agreement with the theoretical expectations,
which predict that almost all of the released gravitational
energy is radiated in all neutrino and anti-neutrino flavours.
Significant conversion of $\bar\nu_e$'s into a sterile neutrino 
would be in conflict with this evidence. 
We can just apply this consideration to constrain the neutrino mixing and 
mass difference. 

We plot in Fig. 6 three contours of the $\bar\nu_e$ 
survival probability ${P}$ for the $\bar \nu_e \rightarrow \bar\nu_s$ 
conversion, in the ($\delta m^2,\sin^2 2 \theta$) parameter space. 
The upper line is for $P = 0.1$,  the lower one is for $P = 0.7$ and 
that in the middle corresponds to $P=0.5$.
If we assume that the successful observation of the SN1987A signal 
implies that at least 50 \% of the expected $\bar\nu_e$ signal has 
been detected,  one can conclude that all the portion above the 
contour of $P=0.5$ 
is ruled out.  For $\delta m^2 = 10^4$ eV$^2$, the range 
$\sin^2 2 \theta > 5 \times  10^{-6}$ is excluded whereas for 
$\delta m^2 = 1\div 10$ eV$^2$ the non-adiabatic character 
of the conversion implies a much looser bound, 
$\sin^2 2 \theta \geq  10^{-1}$. Our results are again in 
qualitative agreement with those in \cite{MS1,SS}. 


Now we would like just to briefly comment on the relevance 
of the large volume detectors, such as Superkamiokande \cite{superkam} 
or SNO \cite{sno} , aimed to detect supernova neutrinos.   
A galactic supernova event would produce e.g. about 5000  events 
through $\bar{\nu}_e p$ reaction in Superkamiokande detector. Such a 
huge statistics may not only allow to determine the neutrino 
flux with good accuracy but also may provide the necessary sensitivity  
to measure e.g. the neutrino energy spectrum. 
The resonant conversion between electron neutrinos and sterile 
neutrinos may show up as a deficit of neutrino events, distortion 
of spectra, time dependent flux, etc.
The absence of a deficit in the expected number of events can be 
used to further constrain the neutrino parameters. In Superkamiokande 
detector a number of $\bar{\nu}_e$ events not larger than 2500 would 
disfavour all the region above the iso-contour of $P=0.5$ in Fig. 6.
However the observation of a possible deficit on electron neutrinos 
or anti-neutrinos is not in itself a distinguishable signal for any 
specific conversion mechanism.
Moreover, the precision of any conclusions that can be drawn from these
considerations is limited by the uncertainties in the theoretical 
neutrino fluxes. 

A better signal is the distortion of the neutrino energy spectrum 
that would arise from the neutrino conversion. The energy dependence 
of the adiabaticity condition may already cause a mild distortion, 
typically neutrinos of lower energies have larger conversion probabilities. 
That is, however, a general  feature of the resonant  conversion
irrespective of the specific neutrino channel. However, we can 
envisage a possible distortion caused by the feedback effect. 
As discussed in Sec. 3, the feedback effect may cause a 30 \% reduction
of neutrino events in the low-energy portion of the neutrino spectrum 
(with a sharp cutoff energy close to 10 MeV). In other words, there
will be a tendency for neutrinos above this energy to be blocked from 
converting. 
In order for this to happen, the following conditions are to be satisfied:
1) the first resonances (for electron neutrinos and anti-neutrinos) must 
be non-adiabatic i.e. for $\delta m^2 < 10^2 \mbox{ eV}^2 $, 
$ \sin^2 2\theta < 10^{-2} $; 
2) the last resonance should be in the region where the electron fraction 
is determined by the neutrino absorption reactions, i.e. for 
$\delta m^2 > 1 \mbox{ eV}^2 $; 
3) the last resonance should be adiabatic.
A more detailed and careful analysis will be given elsewhere\cite{future}.    

\subsubsection{Implications for $r$-process Nucleosynthesis}


The implications of resonant neutrino conversions into active 
neutrinos for the supernova nucleosynthesis have been recently 
investigated in a number of papers \cite{QF,massless}. 

The  $r$-process is responsible for synthesising about half of 
the heavy elements with mass number $A>70$ in nature.
It has been proposed that the $r$-process occurs in the region above the
neutrinosphere in supernovae when significant neutrino fluxes are still
coming from the neutron star \cite{Woosley}. A necessary condition 
required for the $r$-process is $Y_e<0.5$ in the nucleosynthesis region. 
The $Y_e$ value at large radii above the neutrinosphere, where
the $r$-process nucleosynthesis takes place, is determined, as we 
have discussed in Sec. 3, only by the neutrino absorption rates
$\lambda_{\nu_en}$ and $\lambda_{\bar\nu_ep}$.  
Therefore, $Y_e$ in the nucleosynthesis region is approximately given
by the \eq{yelater}. We have also learnt that the presence of neutrino 
conversions into a sterile state can affect the corresponding $Y_e$.  
Thereby, in the nucleosynthesis region we can write $Y_e$ as follows 
\begin{equation}
\label{ye_new}
Y_e \approx {1\over 1+P_{\bar\nu}{\VEV{E_{\bar\nu_e}}}/
P_{\nu}\VEV{E_{\nu_e}}},
\end{equation}
As we already noted the $\bar\nu_e\rightarrow\bar\nu_s$ conversion 
leads to a reduction of $\bar\nu_e$ luminosity and hence to an increase 
of $Y_e$, whereas the $\nu_e\rightarrow\nu_s$ conversion acts in the 
opposite way. Depending on the $\delta m^2$ range, one channel dominates 
over the other one. For $\delta m^2 \geq 10^2$ eV$^2$, only 
$\bar\nu_e\rightarrow\bar\nu_s$ can occur which increases $Y_e$ 
with respect to the case with no anti-neutrino conversion.  
For smaller values of $\delta m^2$ there is an interplay of both 
conversions $\bar\nu_e\rightarrow\bar\nu_s$ and $\nu_e\rightarrow\nu_s$ 
which can make $Y_e < 0.4$, hence enhancing the $r$-process.  

Properly averaging the neutrino absorption rates over the neutrino 
Fermi distribution, we have calculated the electron abundance $Y_e$ 
at the site where the heavy elements nucleosynthesis should take place
as a function of $(\delta m^2, \sin^2 2\theta)$. In Fig. 7 we present 
our result.
 
For a successful $r$-process, the region above $Y_e > 0.5$ is ruled out.  
For $\delta m^2 \geq 10^2$ eV$^2$ and  $\sin^2 2\theta >
2 \times  10^{-6}-10^{-5}$ only the $\bar{\nu}_e \rightarrow 
\bar{\nu}_s$ channel undergoes the conversion. 
As we have discussed in Sec. 4, this 
region is also disfavoured by the neutrino re-heating consideration 
(Fig. 5). 
In correspondence of the corner delimited 
by   50 eV$^2 \leq \delta m^2 \leq 10^2$ eV$^2$ 
and  $\sin^2 2\theta > 10^{-3}$ the conversions take place 
in both channels $\bar{\nu}_e \rightarrow 
\bar{\nu}_s$  and ${\nu}_e \leftrightarrow {\nu}_s$ and are adiabatic enough. 
In this case the significant 
suppression of the $\bar\nu_e$ flux and the  recovery of the
original $\nu_e$ flux at $r_3$ would lead to $Y_e > 0.5 - 1$. 

On the other hand we find that the supernova nucleosynthesis
could be enhanced in  the region enclosed by the dotted contour 
$Y_e=0.4$, delimited by 
$\delta m^2 \leq 10^2$ eV$^2$ and $\sin^2 2\theta \geq 10^{-5}$. 
Inside the contour of $Y_e=0.33$,  $Y_e$ gets stabilised to 
$~$ 1/3 due to the feedback effect (in the absence 
of the feedback  $Y_e$ would be lower than 1/3 - see 
the discussion in  Sec. 3.2).  
We see that the mass range more promising 
for the neutrino hot dark matter scenario, $\delta m^2\leq 10$ eV$^2$, 
is favourable for the $r$-process nucleosynthesis and it is neither 
in conflict with the 
re-heating process (see Fig. 5) nor with SN1987A observations  (see Fig. 6).

\section{Discussion}

In this paper we have investigated the effect of resonant conversions of 
$\nu_e$ or $\bar{\nu}_e$ into sterile neutrinos in the region above the hot 
proto-neutron star in type II supernova. 
For cosmologically interesting mass values (1-100 eV) and mixing angle 
$\sin^2 2 \theta \gsim 10^{-7} \div 10^{-5}$, both $\nu_e$ and $\bar{\nu}_e$ 
could be converted into $\nu_s$ and $\bar{\nu}_s$ (respectively) in the region
outside neutrinosphere due to the non-monotonic behaviour of the effective 
matter potential. 
Such conversion could lead to the depletion of $\nu_e$ and $\bar{\nu}_e$ 
fluxes, resulting in a suppression of the neutrino re-heating behind the 
stalled shock and of the expected $\bar{\nu}_e$ signal in terrestrial 
detectors.
On the basis of these arguments we have derived constraints the 
neutrino mass and mixing parameters. We have found that for 
$\delta m^2 \sin^2 2 \theta \gsim 10^{-3}\mbox{eV}^2$, the energy 
deposition by ${\nu}_e$ and $\bar{\nu}_e$ absorption reactions during 
the shock re-heating epoch ($t<1$ s after the bounce) could be significantly 
decreased. This is not welcome for the delayed explosion scenario which 
relies on the revival of the shock by neutrino re-heating. 

The successful observation of the SN1987A $\bar{\nu}_e$ signal 
in the IMB and Kamiokande detector can also be used in order
to rule out a similar parameter range. Indeed, requiring that 
the total $\bar{\nu}_e$ flux during the thermal neutrino emission 
epoch ($t\sim 1-10$ s p.b.) should not be significantly 
depleted by $\bar{\nu}_e \rightarrow\bar{\nu}_s$ conversion
enables us to rule out the range
$\delta m^2 \sin^2 2 \theta  \gsim 10^{-1}\mbox{eV}^2$.

We further point out that depending on which conversion 
channel, $\nu_e\rightarrow \nu_s$ or $\bar{\nu}_e \rightarrow\bar{\nu}_s$ 
is dominant, $r$-process nucleosynthesis, which might take place in 
neutrino-heated supernova ejects at $1<t\lsim 20$ s p.b., 
could either be suppressed or enhanced. 
For the parameter range $\delta m^2 \sin^2 2 \theta  \gsim 10^{-1}\mbox{eV}^2$
where $\bar{\nu}_e\rightarrow \bar{\nu_s}$ conversion is dominant, $Y_e$ 
at the nucleosynthesis site could become larger than 0.5 and hence 
$r$-process would be prevented, leading to the exclusion of this range 
of mass and mixing.
On the other hand, $r$-process nucleosynthesis could be enhanced due 
to the decrease of $Y_e$ down to the minimum value 1/3. This is due to
the fact that the $\nu_e\rightarrow \nu_s$ conversion is dominant 
if the parameters are in the region $\delta m^2 \lsim 100 \mbox{ eV}^2$ 
and $\sin^2 2 \theta \gsim 10^{-5}$.

We have also discussed that, in contrast to the usual resonant conversion 
among active neutrinos (such as $\nu_e\leftrightarrow \nu_\mu$, $\nu_\tau$), 
the decrease or increase of $Y_e$ due to $\nu_e\rightarrow \nu_s$ or 
$\bar{\nu}_e \rightarrow\bar{\nu}_s$ conversion could be important 
in the estimation of the conversion probabilities.
Indeed, the conversion could be suppressed when $Y_e$ reaches a value 
close to 1/3. Such effect may take place both for $\nu_e \ra \nu_s$ 
or $\bar{\nu}_e \rightarrow\bar{\nu}_s$ resonant conversions. 
However, such feedback effect should not be operative when 
the conversion occurs in the region where the electron or positron 
capture reaction is dominant over the neutrino absorption reaction,  
or if both conversions ($\nu_e$ and $\bar{\nu}_e$ channels) occur in 
the same region.

Finally, we wish to remark on the importance of the supernova
constraints on active sterile neutrino conversions we have derived 
here. Most high-energy particle physics experiments are insensitive 
to the possible existence of sterile neutrinos since these do not 
couple to the electroweak currents. However, their admixture in
the charged current weak interaction could show up in reactor 
neutrino disappearance searches. 
The laboratory limits on the mixing between electron neutrinos 
and sterile neutrinos are quite weak for this mass range. From reactor 
experiments the bound is weaker than $\sin^2 2 \theta \lsim 0.01$ 
\cite{reactor},
far weaker than the supernova limits we derive.  In contrast, for large mixing
the reactor limit on $\delta m^2 \lsim 10^{-2}$ eV$^2$  is stronger than
the corresponding supernova limits. 

A much stronger though also more uncertain argument to constrain 
active-sterile neutrino conversions comes from big-bang cosmological 
nucleosynthesis. Assuming that the number of effective neutrino species
is bounded to be less than 4, one has \cite{BBN}:
\begin{eqnarray}
\label{BBNL}
\delta m^2 \, \sin^42\theta &\lsim & 5 \times  10^{-6} \mbox{eV}^2 ~,
~~~~ 
\nu=\nu_e,  \\
\delta m^2\,\sin^42\theta &\lsim & 3 \times  10^{-6} \mbox{eV}^2 ~, ~~~~
\nu=\nu_\mu~, \nu_\tau ~.  
\end{eqnarray}
However the recent discrepant observational determinations of the 
primordial deuterium abundances \cite{dhigh,dlow} may force us to 
revise big-bang nucleosynthesis constraints on nonstandard neutrino 
physics. As a result the assumptions on which the previous limit is 
based are under debate \cite{FV}. It has been argued that $N_\nu>4$ 
is acceptable \cite{KS}, in which case there would be room to bring 
an extra sterile neutrino species into equilibrium with the known 
neutrinos in the early Universe and therefore no constraints on 
active to sterile neutrino oscillation parameters. 
Therefore, from this point of view, the restrictions on active-sterile 
neutrino oscillation parameters obtained from supernova theory and 
observations become quite relevant.

\vspace{0.7cm}

\centerline{\bf Acknowledgement}

We thank Y.-Z. Qian for stressing to us the importance of feedback 
effects in supernovae and for a helpful discussion. This work has been 
supported by DGICYT under Grant N. PB95-1077, by a joint CICYT-INFN 
grant, and by the TMR network ERBFMRXCT960090 of the European Union. 
H. N. has been supported by Ministerio de Educaci\'on y Ciencia and 
A. R. by the Human Capital and Mobility Program under Grant No. 
ERBCHBI CT-941592 and by the Grupo Teorico de Altas Energias at 
Instituto Superior Tecnico (Lisboa). A. R. thanks the CERN Theory 
Division for hospitality during part of this work.  H. N. thanks 
the Institute for Nuclear Theory at the University of Washington 
for its hospitality during the part of this work.


\newpage
\bef
\centerline{\protect\hbox{
\psfig{file=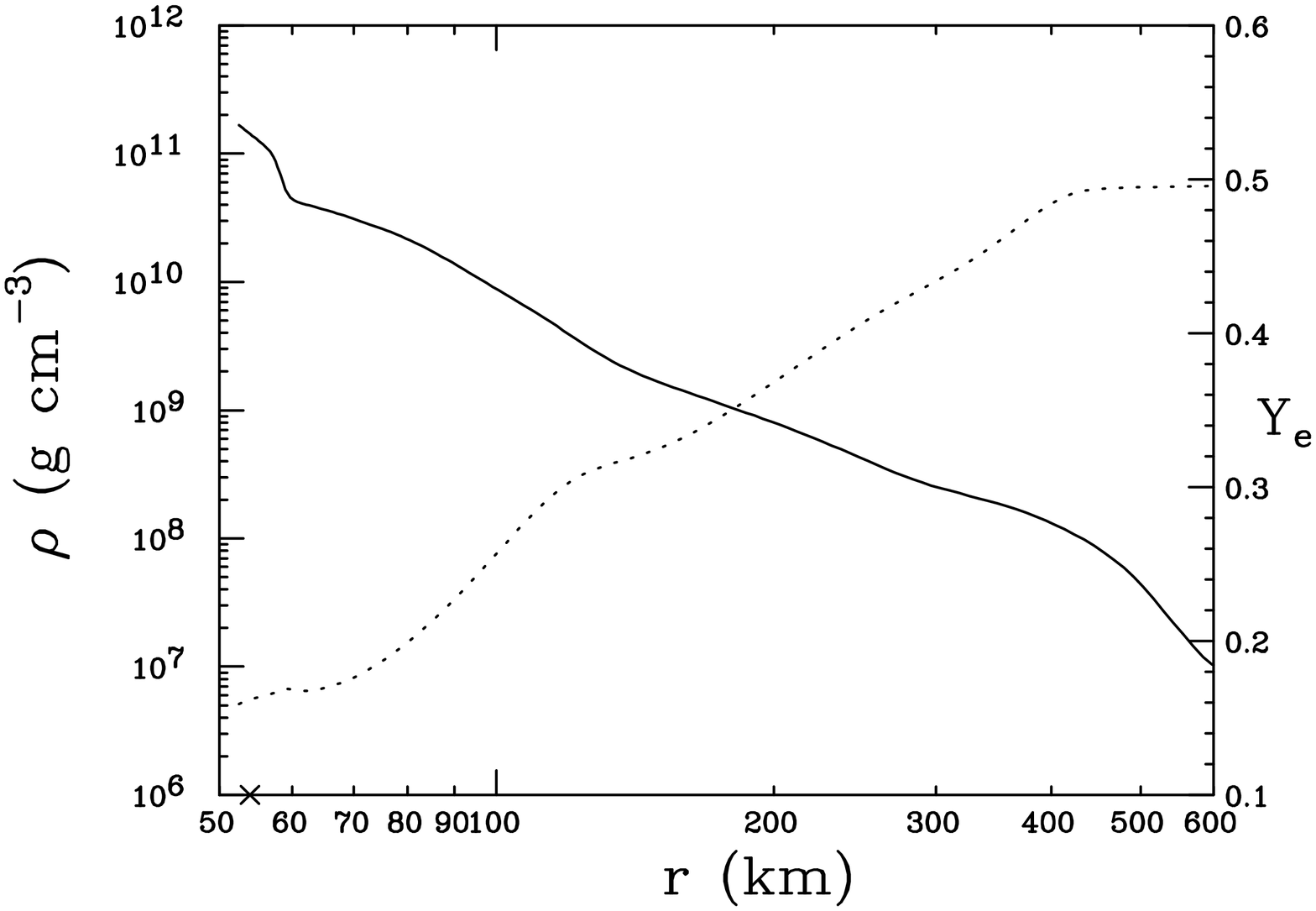,height=13.0cm,width=16.0cm,angle=90}
}}
\noindent Fig. 1a: 
Typical matter density (solid line) and $Y_e$ (dotted line) 
profiles versus the radial distance from the center of the star, in 
Wilson's numerical supernova model at $t=0.15$ s after the core bounce. 
The diagonal cross indicates the position of the surface of 
the neutrinosphere.

\eef
\bef
\centerline{\protect\hbox{
\psfig{file=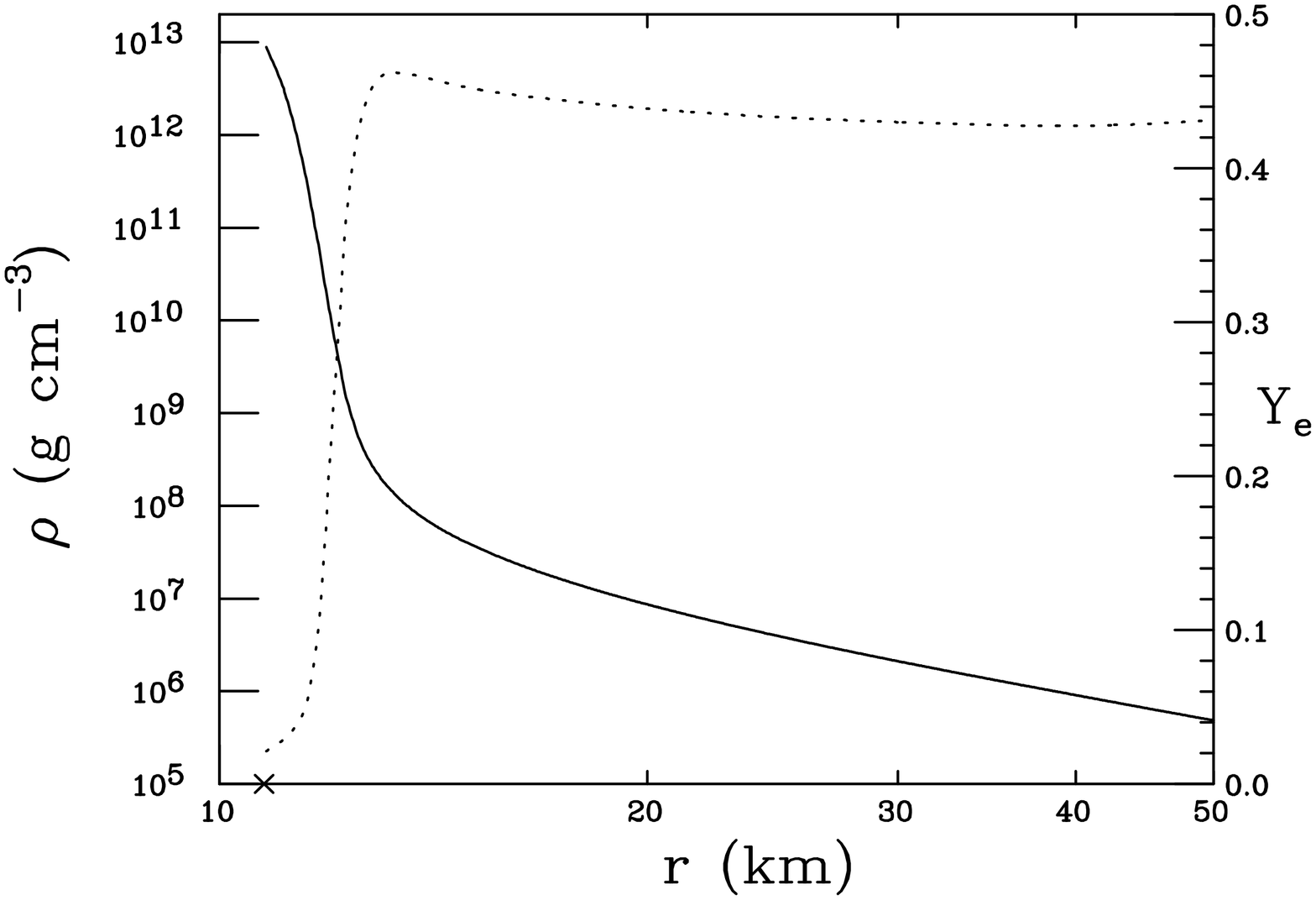,height=13.0cm,width=16.0cm,angle=90}
}}
\noindent Fig. 1b: 
Same as in Fig. 1a but for $t\sim 6$ s after the core bounce.

\eef
\bef
\centerline{\protect\hbox{
\psfig{file=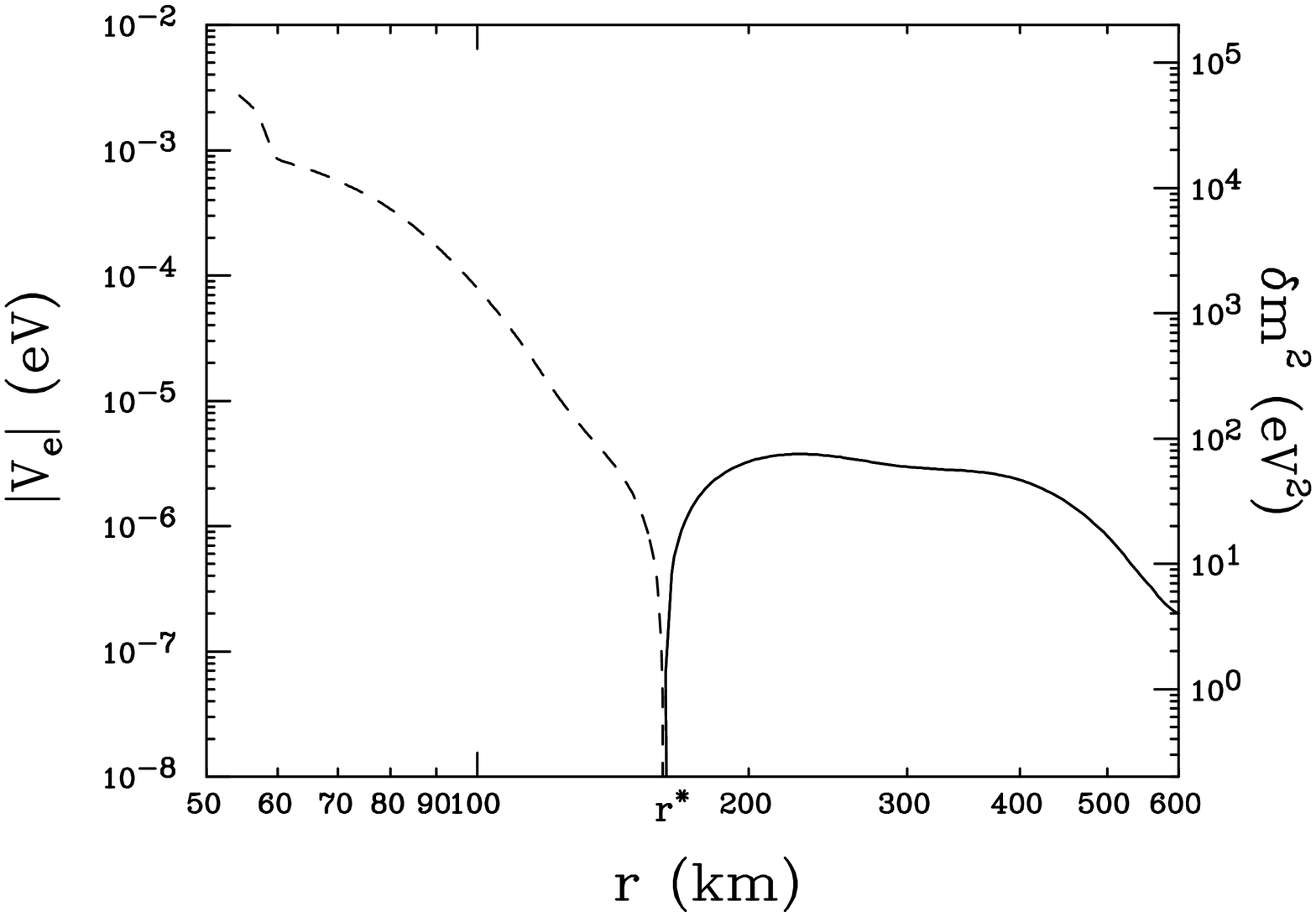,height=13.0cm,width=16.0cm,angle=90}
}}
\noindent Fig. 2a: 
The modulus of the matter potential $V_e$ in \eq{potential} 
versus the radial distance $r$ from the center of the star.
This is obtained using the matter density and $Y_e$ profiles 
in Fig. 1a, at $t<1$ s after the core bounce. The solid and 
dashed lines correspond to positive and negative potential, 
respectively, and the position where $V_e=0$ is denoted by 
$r^*$. We also indicate, in the right ordinate, the $\delta m^2$ 
values for which a $E=10$ MeV neutrino undergoes resonant 
conversion, for the corresponding value of $|V_e|$ on the 
left ordinate (small mixing angle is understood).
\eef
\bef
\centerline{\protect\hbox{
\psfig{file=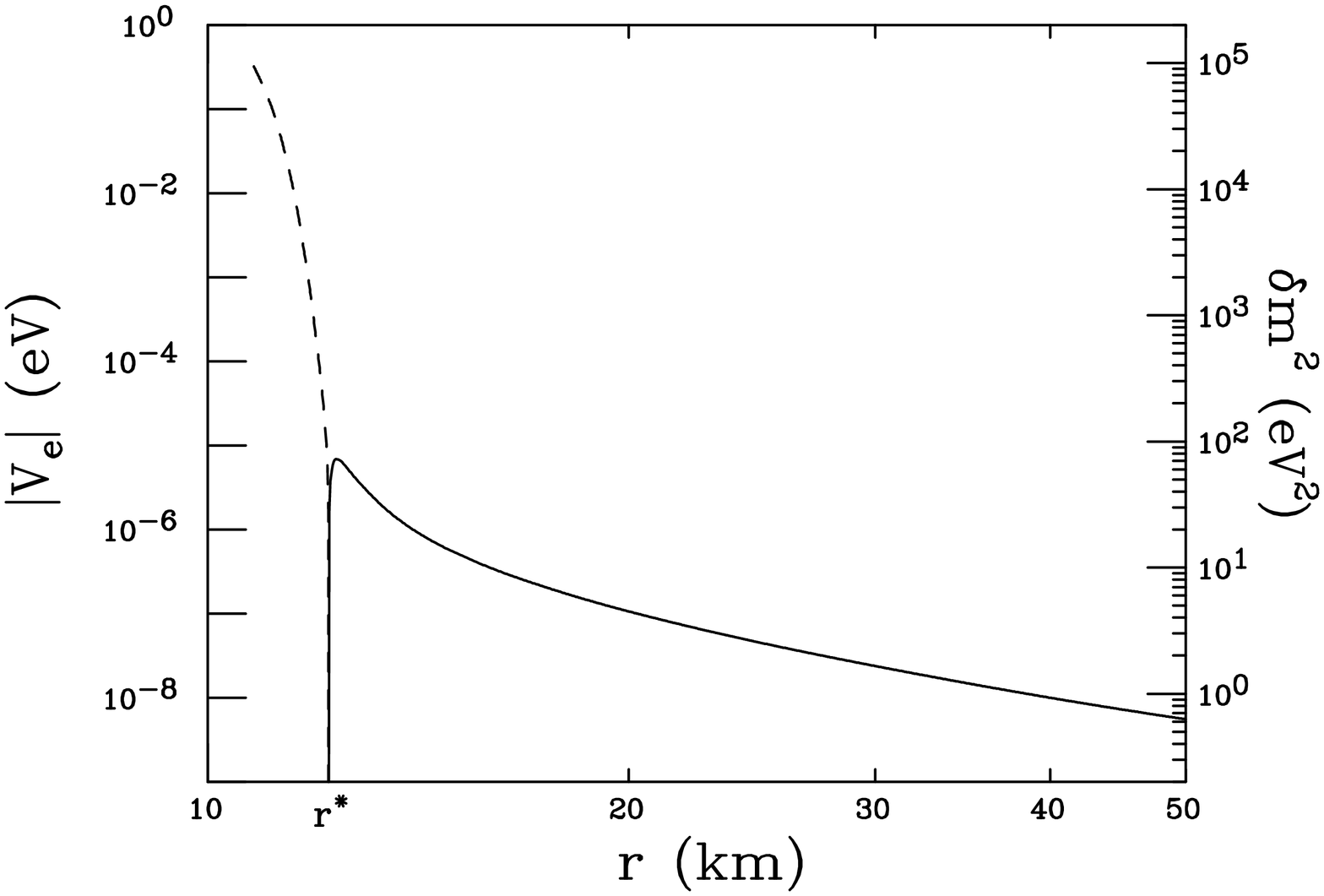,height=13.0cm,width=16.0cm,angle=90}
}}
\noindent Fig. 2b: 
Same as in Fig. 2a but for $t>1$ s after the core bounce.

\eef

\newpage
\bef
\centerline{\protect\hbox{
\psfig{file=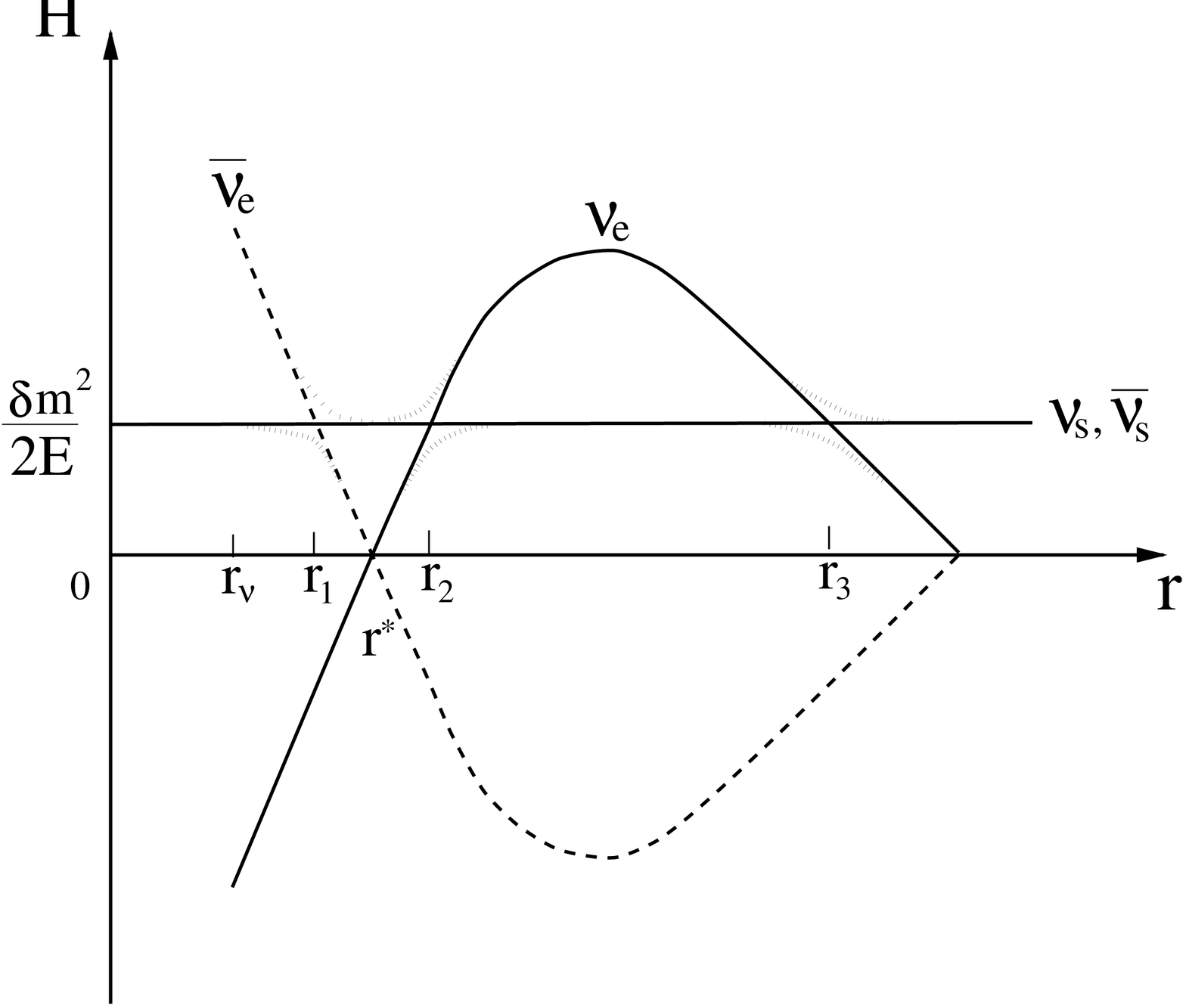,height=14.0cm,width=15.0cm,angle=-90}
}}
\vglue 0.3cm
\noindent Fig. 3: Schematic figure illustrating the level crossings
for $\nu_e-\nu_s$ and $\bar\nu_e-\bar\nu_s$ system. The energy levels 
of $\nu_e$ (solid line), $\bar\nu_e$ (dashed line) and $\nu_s$, 
$\bar\nu_s$ (horizontal line) are given as functions of the stellar 
radius. The positions where $\bar\nu_e \rightarrow \bar\nu_s$ and 
$\nu_e\leftrightarrow \nu_s$ resonances occur are 
indicated by $r_1$, $r_2$ and $r_3$, respectively. 

\eef

\bef
\centerline{\protect\hbox{
\psfig{file=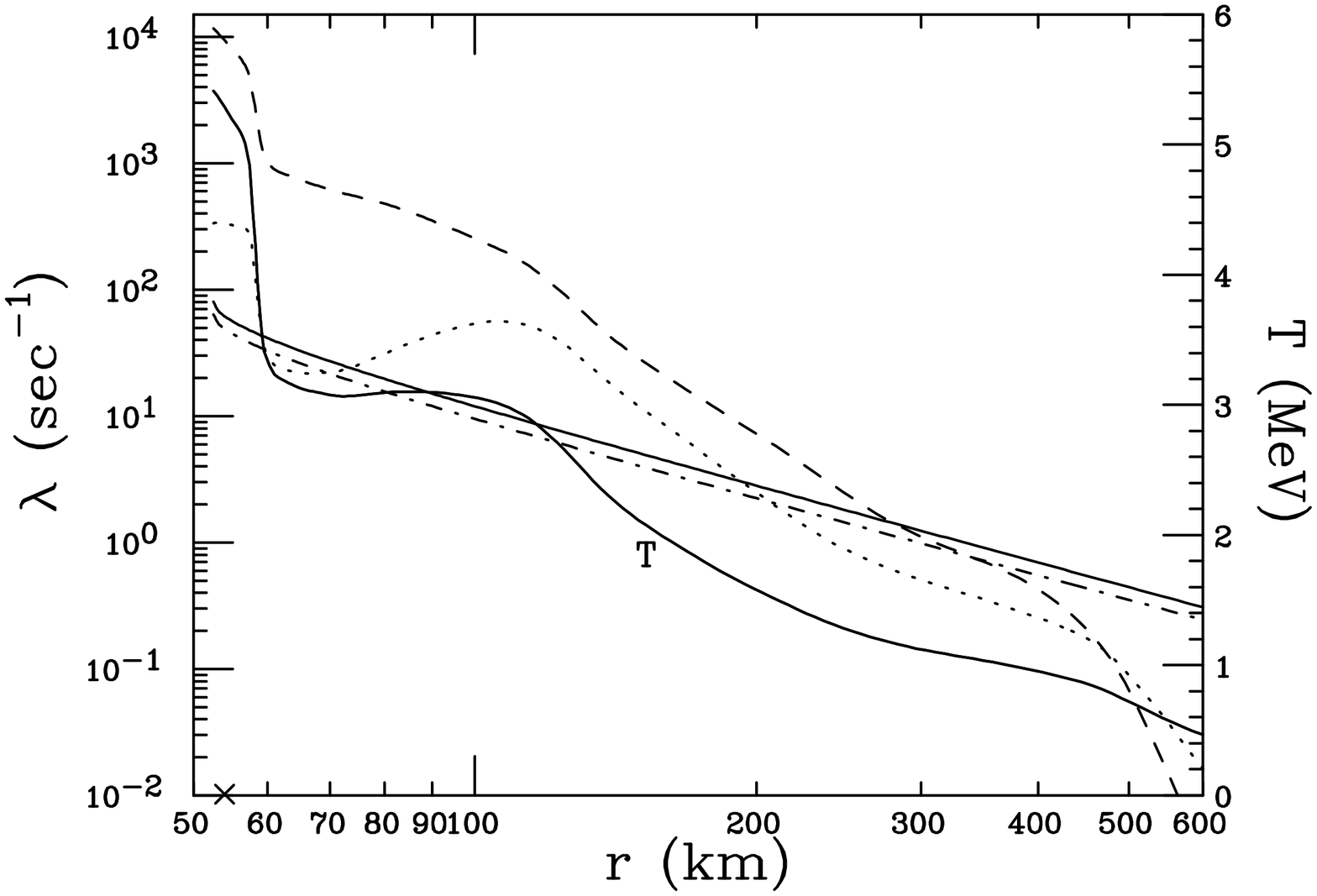,height=13.0cm,width=16.0cm,angle=90}
}}
\noindent Fig. 4a: 
Main neutrino reaction rates (left ordinate) versus the radial distance 
from the stellar center at $t< 1$ s after the core bounce: 
$\nu_e n \ra p e^{-}$ (solid curve),  $\bar\nu_e p\rightarrow n e^{+}$ 
(dot-dashed curve), $e^{-} p\rightarrow \nu_e  n$ (dashed curve) and 
$e^{+} n\rightarrow \bar\nu_e  p$ (dotted  curve). The temperature 
profile (right ordinate) is also shown by the solid line (labelled $T$). 
We assume a neutrino (and anti-neutrino) luminosity of $L_\nu=10^{52}$ ergs/s.

\eef
\bef
\centerline{\protect\hbox{
\psfig{file=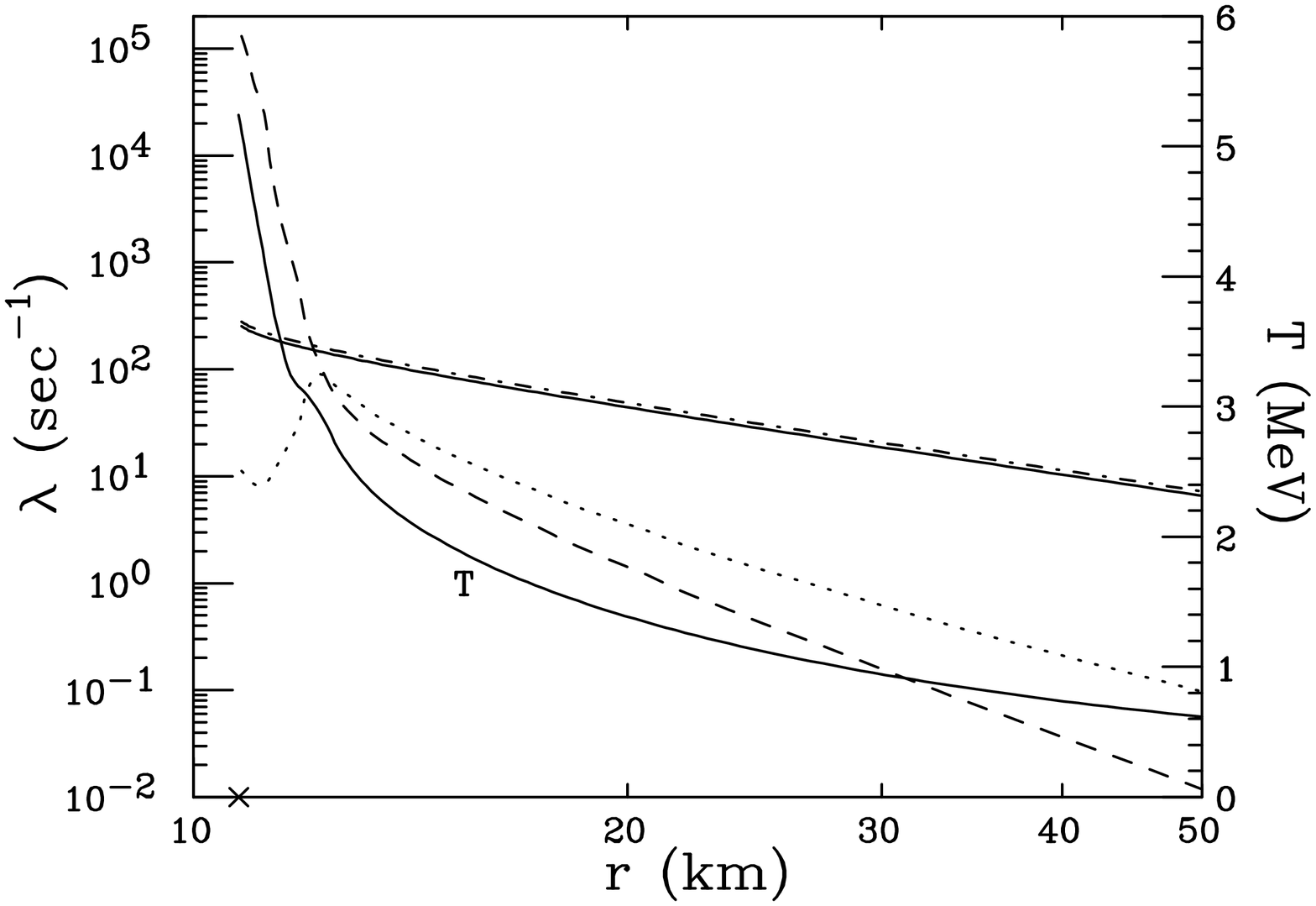,height=13.0cm,width=16.0cm,angle=90}
}}
\noindent Fig. 4b: 
Same as in Fig. 4a but for $t>1$ s after the core bounce.

\eef

\newpage
\bef
\centerline{\protect\hbox{
\psfig{file=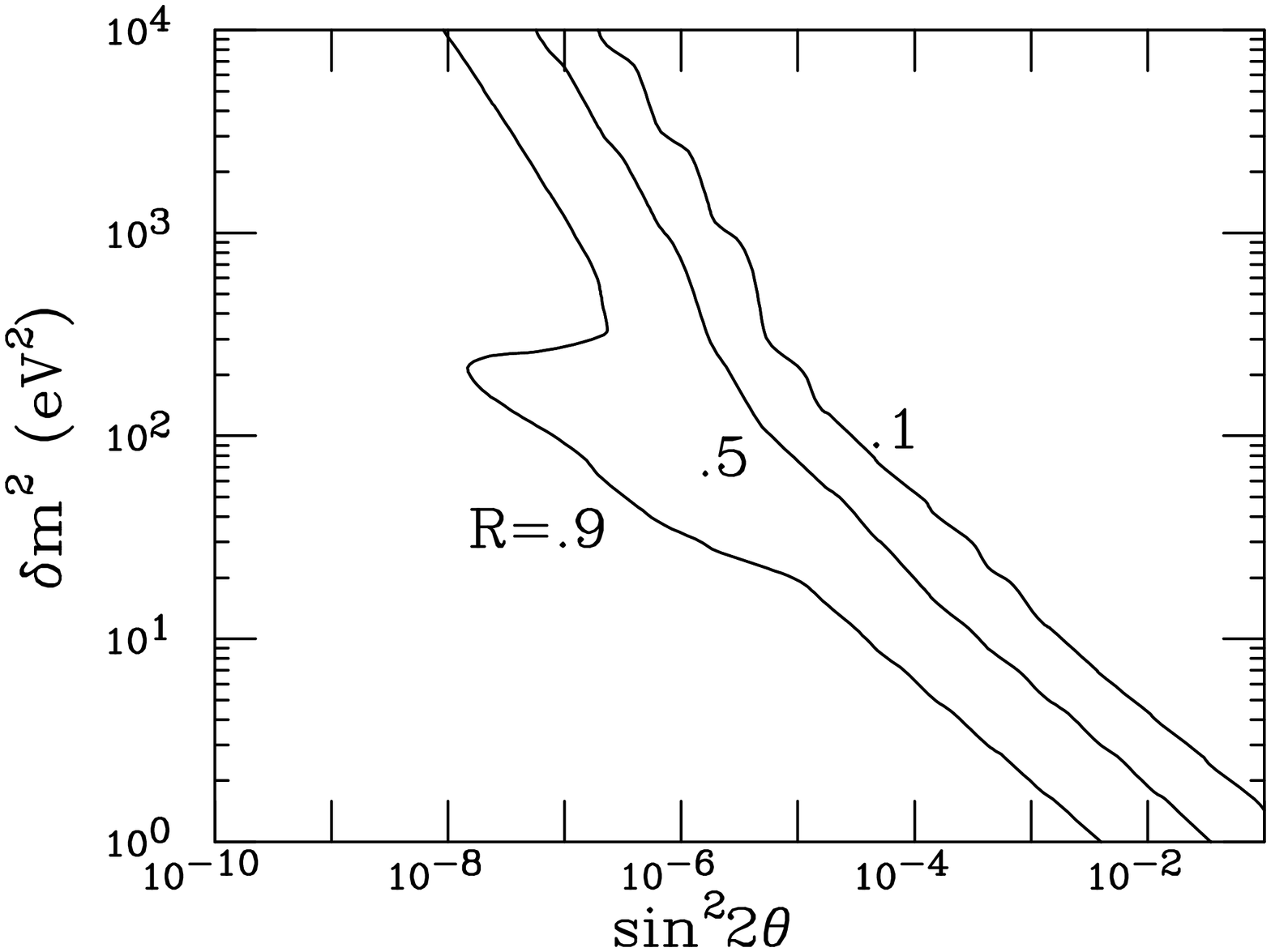,height=13.0cm,width=16.0cm,angle=90}
}}
\noindent Fig. 5: 
Contour plot of the ratio $R$ of the neutrino energy deposition 
behind the shock wave in the presence of conversions into 
sterile neutrinos, versus the case without conversions, 
as defined in eq. (15).

\eef

\bef
\centerline{\protect\hbox{
\psfig{file=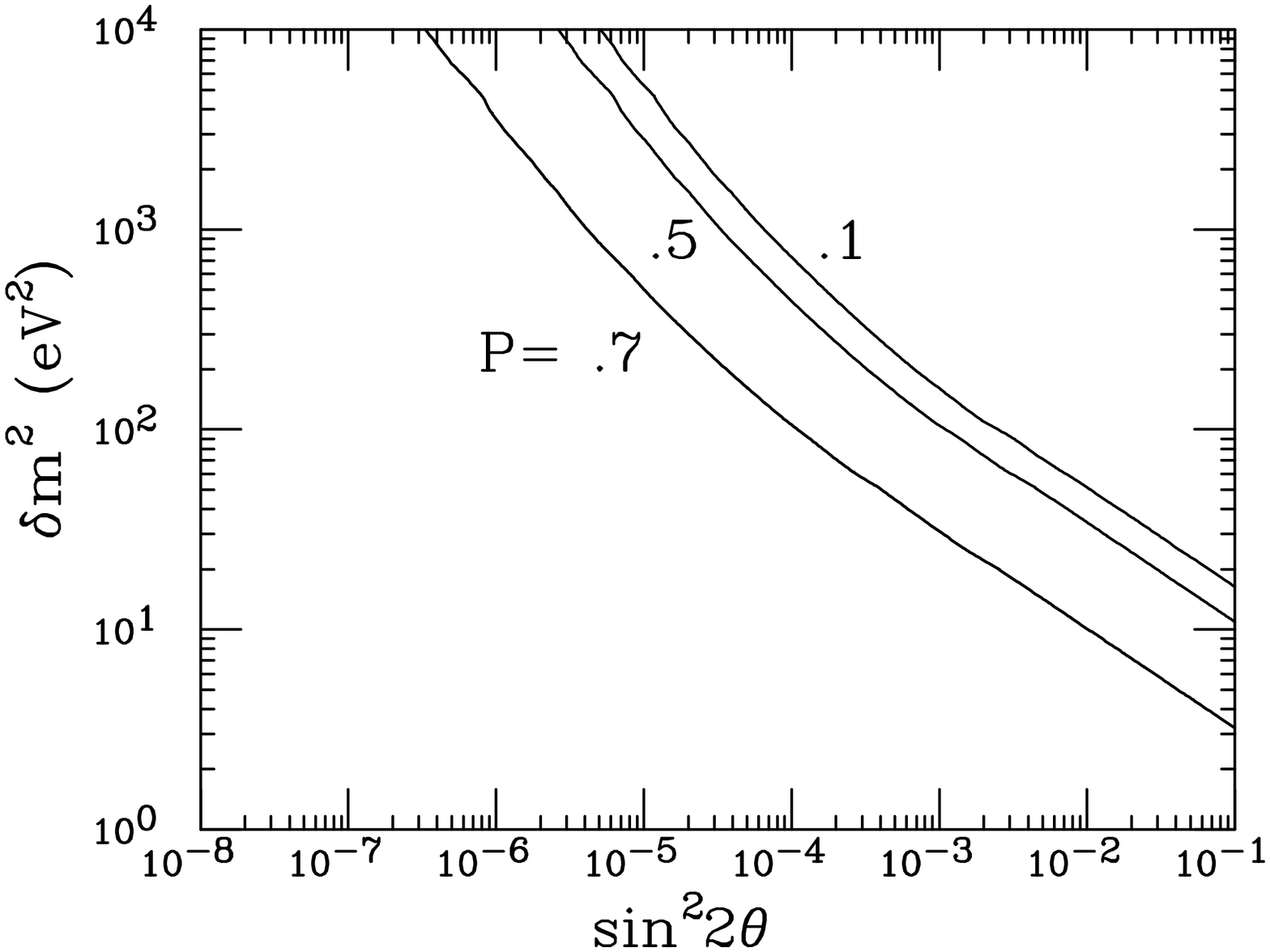,height=13.0cm,width=16.0cm,angle=90}
}}
\noindent Fig. 6: 
Contour plots of the survival probability $P$ (figures at the curve) 
for the $\bar\nu_e\rightarrow\bar\nu_s$ conversion at $t>1$ s p.b. 
The region to the right of the curves can be excluded by the observation 
of the SN1987A $\bar\nu_e$ signal.

\eef
\bef
\centerline{\protect\hbox{
\psfig{file=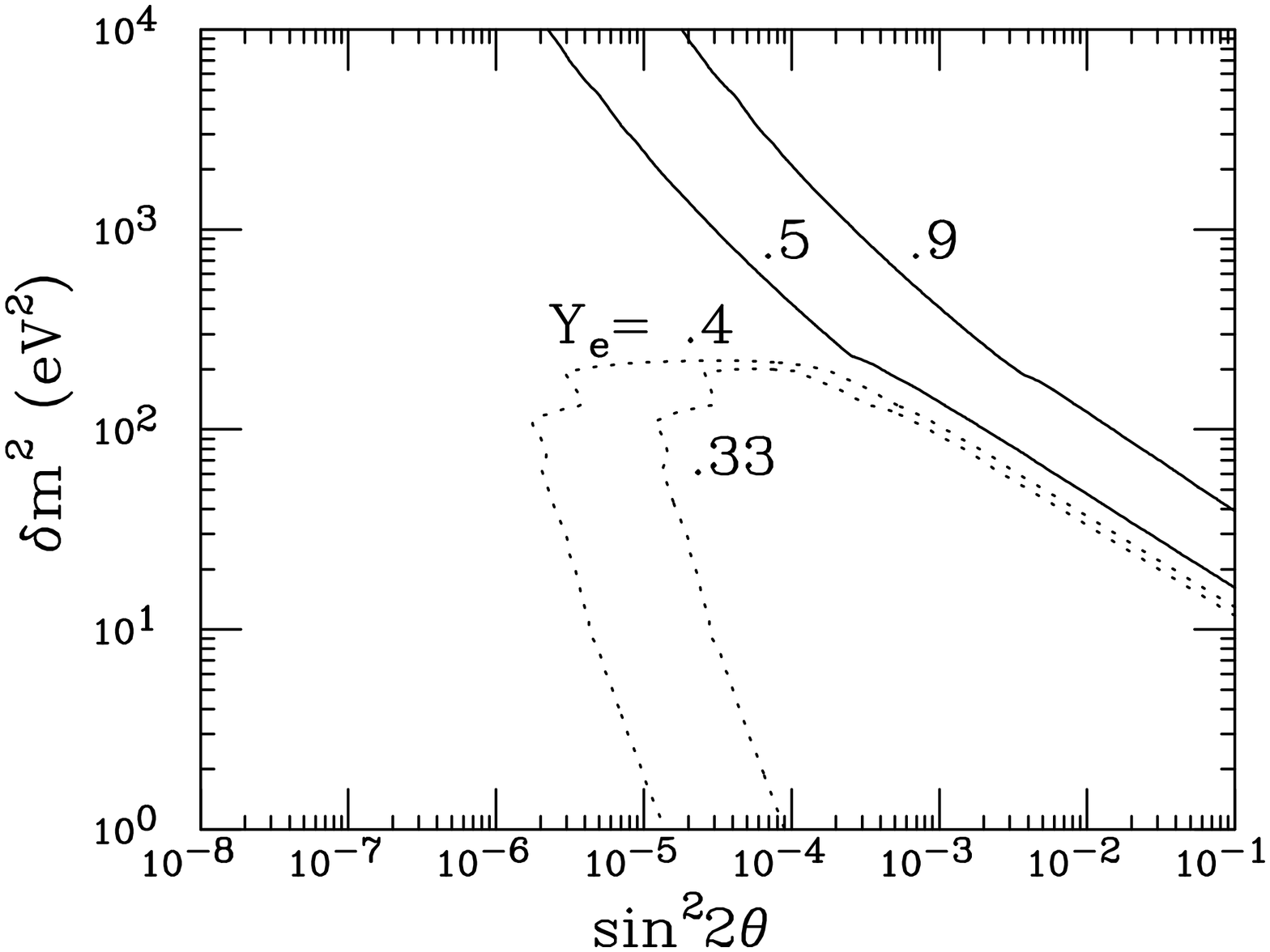,height=13.0cm,width=16.0cm,angle=90}
}}
\noindent Fig. 7: 
Contour plots for the electron concentration $Y_e$ (figures at the curves)
taking into account $\bar\nu_e\rightarrow\bar\nu_s$ and 
$\nu_e\rightarrow\nu_s$ conversions at $t>1$ s p.b. The region to the 
right of the solid line labelled 0.5 is ruled out by the condition
$Y_e<0.5$ necessary for $r$-process nucleosynthesis to occur. For the 
parameter region inside the $Y_e= 0.4$ dotted contour $r$-process 
nucleosynthesis can be enhanced.

\eef

\end{document}